# Demonstration of broadband topological slow light

Fujia Chen[1,2,3,#], Haoran Xue[4,5,#], Yuang Pan[1,2,3], Maoren Wang[6], Yuanhang Hu[6], Li Zhang[1,2,3], Qiaolu Chen[1,2,3], Song Han[1,2,3], Gui-geng Liu[4,5], Zhen Gao[7], Peiheng Zhou[6], Wenyan Yin[1], Hongsheng Chen[1,2,3,*], Baile Zhang[4,5], Yihao Yang[1,2,3,*]

[1]Interdisciplinary Center for Quantum Information, State Key Lab. of Modern Optical Instrumentation, College of Information Science and Electronic Engineering, Zhejiang University, Hangzhou 310027, China.

[2]ZJU-Hangzhou Global Science and Technology Innovation Center, Key Lab. of Advanced Micro/Nano Electronic Devices & Smart Systems of Zhejiang, ZJU-UIUC Institute, Zhejiang University, Hangzhou 310027, China.

[3]Jinhua Institute of Zhejiang University, Zhejiang University, Jinhua 321099, China.

[4]Division of Physics and Applied Physics, School of Physical and Mathematical Sciences, Nanyang Technological University, 21 Nanyang Link, Singapore 637371, Singapore.

[5]Centre for Disruptive Photonic Technologies, The Photonics Institute, Nanyang Technological University, 50 Nanyang Avenue, Singapore 639798, Singapore.

[6]National Engineering Research Center of Electromagnetic Radiation Control Materials, State Key Laboratory of Electronic Thin Film and Integrated Devices, University of Electronic Science and Technology of China, 610054 Chengdu, China.

[7]Department of Electrical and Electronic Engineering, Southern University of Science and Technology, Shenzhen 518055, China.

[#]These authors contributed equally.

*E-mail: yangyihao@zju.edu.cn (Y. Y.); hansomchen@zju.edu.cn (H. C.);

# Abstract

Slow-light devices are able to significantly enhance light-matter interaction due to the reduced group velocity of light, but a very low group velocity is usually achieved in a narrow bandwidth, accompanied by extreme sensitivity to imperfections that causes increased disorder-induced attenuation. Recent theories have suggested an ideal solution to this problem—unidirectional chiral photonic states, previously discovered in structures known as photonic topological insulators, not only resist backscattering from imperfections but can also be slowed down in the entire topological bandgap with multiple windings in the Brillouin zone. Here, we report on the experimental demonstration of broadband topological slow light in a photonic topological insulator. When coupled with periodic resonators that form flat bands, the chiral photonic states can wind many times around the Brillouin zone, achieving an ultra-low group velocity in the entire topological bandgap. This demonstration extends the scope of topological photonics into slow light engineering, and opens a unique avenue in the dispersion manipulation of chiral photonic states.

It is well-known that when passing through certain media, light can be slowed down with a group velocity remarkably lower than that in vacuum[1-3]. This slow-light effect can dramatically enhance light-matter interaction and significantly minimise photonic device footprints, enabling numerous photonic applications in switching, storage, buffers, and quantum optics[4-8]. In this context, tremendous efforts have been made, based on conventional photonic crystal waveguides[3,9-11] or coupled-resonator optical waveguides[12-14]. For example, in the vicinity of a photonic band edge, the group velocity of the waveguide modes can be extremely low (see Fig. 1a). However, these waveguide modes are usually quite dispersive and narrowband. Moreover, a very low group velocity usually implies extreme sensitivity to disorders from fabrication imperfections, leading to significant losses arising from backscattering and Anderson localisation[15-18]. This, therefore, has imposed a fundamental constraint on the performance of slow-light devices in practice.

The recently discovered photonic topological insulators can sustain topologically protected chiral edge states (CESs) that are intrinsically immune to backscattering and Anderson localisation[19-37]. The existence of CESs is guaranteed by the nonzero Chern number of the bulk, according to the principle of bulk-edge correspondence[38]. The edge modification in a photonic topological insulator can only change the edge dispersions, but cannot change the number of CESs[33]. The photonic topological insulators are, therefore, an ideal platform to resolve the challenge of disorder-induced scattering losses in slow-light devices[22,23,39,40]. Conventionally, the CESs exhibit a dispersion traversing the bulk bandgap in a single Brillouin zone[22]. In this scenario, the topological slow light can be achieved by engineering the dispersion around a given frequency (see Fig. 1b)[41-43], which, however, implies a narrow operating bandwidth.

To circumvent the aforementioned challenges, i.e., disorder-induced attenuation and narrowband operation, broadband topological slow light through Brillouin zone winding of the CESs has been proposed recently [44-46]. Without sacrificing the operating bandwidth, the group velocity of CESs can be dramatically reduced when their dispersion winds around the Brillouin zone multiple times (see Fig. 1c). Typically, there are two approaches to achieving broadband topological slow light. The first is to modify the nearest-neighbour and next-nearest-neighbour couplings in a topological insulator lattice, whose design has not yet been translated into a realistic photonic setting[44]. The second is to load judiciously designed resonators at the edge termination of a photonic topological insulator[45,46]. As a result, the CESs couple with the flat bands induced by resonance modes, and wind around the Brillouin zone multiple times with a reduced group velocity. So far, both types of Brillouin zone winding for CESs have not been explored experimentally.

Here, we report on the experimental demonstration of broadband topological slow light through Brillouin zone winding. Our work mainly follows the second approach, i.e., loading properly tailored resonators at the edge termination of a photonic topological insulator[45,46]. We experimentally show that by coupling to the resonance-induced flat bands, the CESs can wrap many times around the Brillouin zone. Consequently, their group velocity can be significantly slowed down over a broad bandwidth. Interestingly, the newly-realised hybridised CESs are reminiscent of the polaritons that result from the strong coupling between photons and electric dipoles. Finally, we also demonstrate the robustness of such broadband topological slow light against obstacles and defects, as well as the smooth transport through slow-light regions with different group velocities.

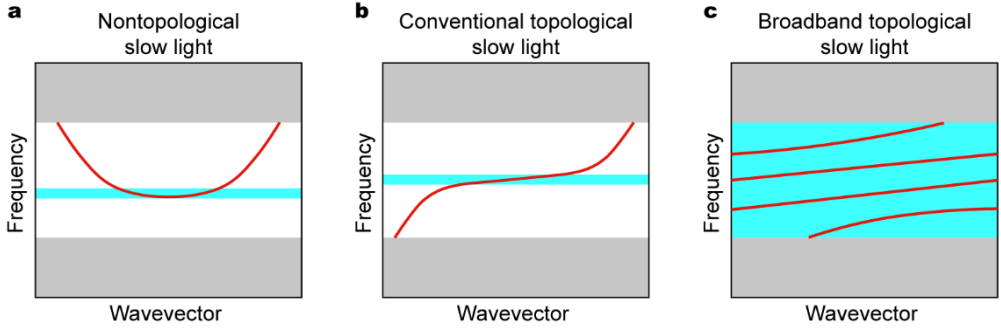

**Fig. 1 | Comparison between different approaches to slow light. a,** Nontopological slow-light systems. They usually suffer from narrow bandwidths and susceptibility to disorders caused by fabrication imperfections. **b,** Conventional topological slow-light systems. In these systems, the CESs generally cross the bandgap within a single Brillouin zone, which can be engineered to exhibit low group velocities in a narrow band. **c,** Broadband topological slow-light systems through Brillouin zone winding. In these systems, the dispersion of CESs can wrap many times around the Brillouin zone, enabling broadband, topologically protected slow light. The blue rectangles represent the effective bandwidth of slow light in three different approaches.

**Experimental setup and schematic.**
To demonstrate the broadband topological slow light through Brillouin zone winding, we design an experimental setup as shown in Fig. 2**a,b**. The photonic topological insulator consists of a square lattice of magnetically-biased yttrium-iron-garnet (YIG) rods (see Method for material parameters) with lattice constant $a$ = 14.6 mm and rod radius $d$ = 4.4 mm. In this work, we focus on the transverse electric (TE) modes with electric fields polarised along the YIG rods. Several YIG rods are enclosed by gratings of metal cylinders to construct photonic cavities that are periodically loaded at the edge of the photonic topological insulator. As the distance between two neighbouring metal cylinders ($\delta = a/3$) is much smaller than the cut-off wavelength ($\lambda_0/2$, with $\lambda_0$ the operating wavelength) for TE modes, the metal gratings behave approximately as metal walls[47](see more details in Supplementary Information). The resonance modes in cavities are determined by the cavity width $\Lambda$ ($\Lambda = 2a$ in our designs, and $\Lambda$ is also the period of the resonator array) and the cavity depth $D$ ($D = 2a$ in Fig. 2**a,b**). The gap width $g$ ($g = 4\delta$ in Fig. 2**a,b**) at the upper side of resonators controls the coupling between the CESs and the resonance modes. All YIG rods are placed in an air-loaded waveguide composed of two parallel copper plates and magnetised by small samarium cobalt (SmCo) permanent magnets locating at the top and bottom of each YIG rod (see more details in Supplementary Information), as shown in Fig. 2**a**. Here, a part of the top cover plate is removed for visualisation. Note that a square array of small holes, with a diameter of 2 mm and a period of $a/3$, are drilled through the top copper plate for the convenience of inserting metal cylinders and probes. These holes have a negligible influence on the wave propagation confined in the waveguide. The red arrow in Fig. 2**a** represents the position of the source antenna. The numerically calculated band structure of the photonic topological insulator is shown in Fig. 2**c**. Under a z-oriented external static magnetic field with B = 0.043 Tesla, there is a topological Chern-type bandgap from 11.5 GHz to 12.9 GHz (indicated by the grey rectangle in Fig. 2**c**) with a gap Chern number $C_{gap}$ = 1, in which the gapless CESs exist (see more details in Supplementary Information).

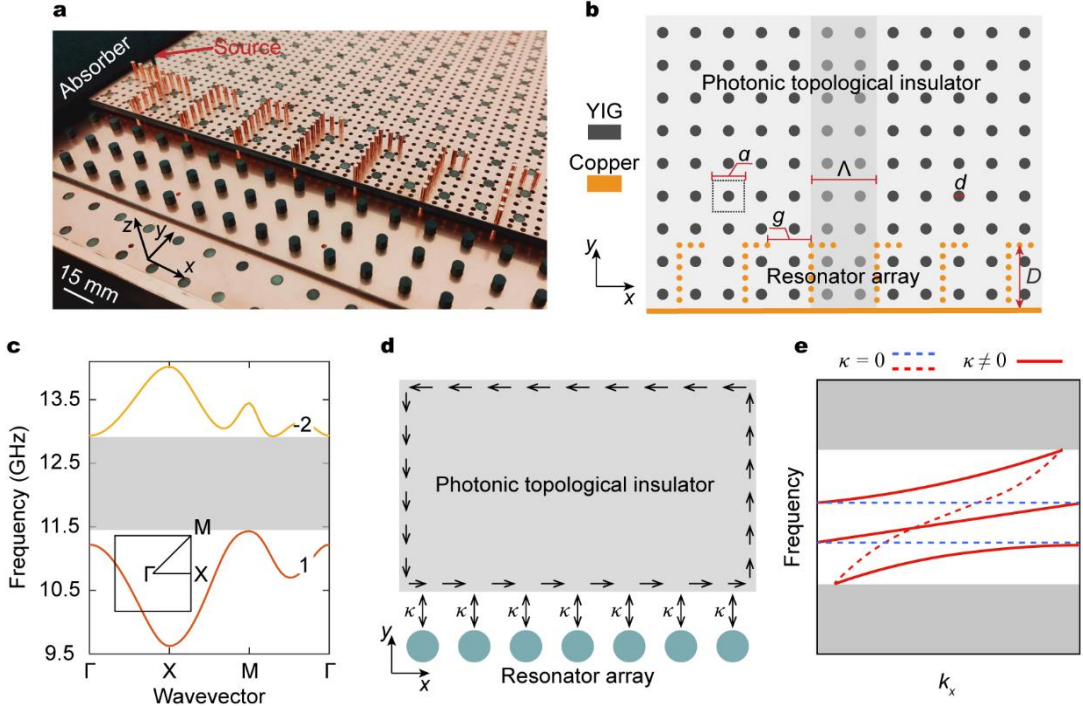

**Fig. 2 | Experimental setup and schematic. a-b,** Photograph and diagram of the experimental setup. The photonic topological insulator with lattice constant $a$ = 14.6 mm is placed in an air-loaded parallel-plate waveguide with a height of 5 mm. A square array of small holes, with a diameter of 2 mm and a period of $a/3$, is drilled through the top copper plate. The red arrow in (**a**) shows the position of the source antenna. The dashed box in (**b**) indicates the unit cell of the photonic topological insulator. The dark grey circles in (**b**) represent YIG cylinders with a diameter $d$ = 4.4 mm. The yellow dots and line represent the copper cylinders and the copper bottom boundary, respectively. Here, $g$, $D$, and $\Lambda$ denote the gap width, the cavity depth, and the cavity width, respectively. **c,** Second and third bands of the gyromagnetic photonic crystal under an external static magnetic field with B = 0.043 Tesla. There is a nontrivial topological bandgap from 11.5 to 12.9 GHz (marked by a grey rectangle) in which CESs exist. The Chern number of each band is labelled. Note that the first band is topologically trivial which is not shown here. The insert shows the Brillouin zone. **d,** Schematic of the mechanism. The coefficient $\kappa$, characterising the reciprocal coupling strength (depicted by double arrows) between the CESs and the resonance modes, is controlled by the gap width $g$. **e,** Dispersions of edge modes in the decoupling (i.e., $\kappa = 0$, the dashed curves) and coupling (i.e., $\kappa \neq 0$, the solid curves) regimes. The blue and red dashed curves represent the flat bands and the CESs,

respectively.

The experimental setup in Fig. 2**a** can be understood from the schematic in Fig. 2**d**. The resonators, periodically arranged at the bottom edge of a photonic topological insulator, are isolated from each other and coupled to the CESs with a coupling strength $\kappa$. When $\kappa = 0$, the photonic topological insulator is equivalently terminated with a perfect electric conductor (PEC) boundary where the CESs must exist, according to the celebrated bulk-edge correspondence. For a typical photonic topological insulator with Chern number $|C|=1$, a single edge dispersion traverses the entire topological bandgap[21,22] (see the dashed red curves in Fig. 2**e**). Due to the decoupling, the crossing points between the chiral edge dispersion and the flat bands are preserved. When $\kappa \neq 0$, the CESs couple to the resonances (i.e., the flat bands denoted by the dashed blue lines in Fig. 2**e**), splitting up the crossing points. The resulting hybridised edge states are still chiral, i.e., unidirectional, and the corresponding dispersion crosses the entire topological bandgap (see the solid red curves in Fig. 2**e**). In comparison with the original CESs, the hybridised CESs exhibit a dispersion wrapping multiple times around the Brillouin zone. Over a broad bandwidth, the group velocity of the hybridised CESs, defined as the slope of the dispersion curve, i.e., $v_g = d\omega/dk$, where $\omega$ and $k$ are the angular frequency and the wavevector, respectively, is significantly slowed down, with the slowdown factor proportional to the number of flat bands available in the topological bandgap.

**Coupling between the CESs and the flat bands.**
We start with the experimental investigation of the coupling process between the CESs and the flat bands. Figure 3**a-f** show the experimentally measured (the background colormap) and numerically calculated (the curves) dispersions of the CESs and the flat bands, when the coupling strength increases, which is positively related to the gap width $g$. When $g$ is very small (e.g., $g = \delta$), the evanescent coupling between the CESs and the flat bands is negligible, and the dispersion of CESs (the black dashed curves) and the flat band (the blue dashed lines) induced by the resonance modes intersect with each other, as shown in Fig. 3**a**. Correspondingly, the CESs and the resonance modes do not hybridise with each other, as shown in Fig. 3**g**. As the coupling strength increases, the CESs are strongly coupled with the resonance modes, forming hybridised CESs (see Fig. 3**g**). Consequently, the crossing points are split up, and the dispersion curves of the CESs and the flat bands connect with each other, to produce a single dispersion curve winding twice around the Brillouin zone in the topological bandgap. Obviously, the hybridised CESs exhibit much lower group velocities but the same operational bandwidth, i.e., the topological bandgap width. Continuously increasing $g$, the coupling strength is so considerable that the flat bands shift, and so do the dispersions of hybridised CESs (see Fig. 3**d-f**). In particular, at $g = 4\delta$, the second flat band shift into the topological bandgap, enabling one additional winding of the hybridised CESs, as shown in Fig. 3**d,g**. In the following, we will focus on the cases with $g = 4\delta$.

Note that the hybridised CESs formed by coupling the propagating CESs and resonance modes are reminiscent of the polaritons that result from the strong coupling between light and dipolar oscillations, such as phonon polaritons and exciton polaritons[48,49]. Our newly-realised hybridised CESs are, therefore, a sort of one-way topological polaritons. Besides, in stark contrast to the conventional polaritons with two separate dispersion branches, our hybridised CESs have a single dispersion curve, as the two dispersion branches reconnect with each other at the boundary of the Brillouin zone.

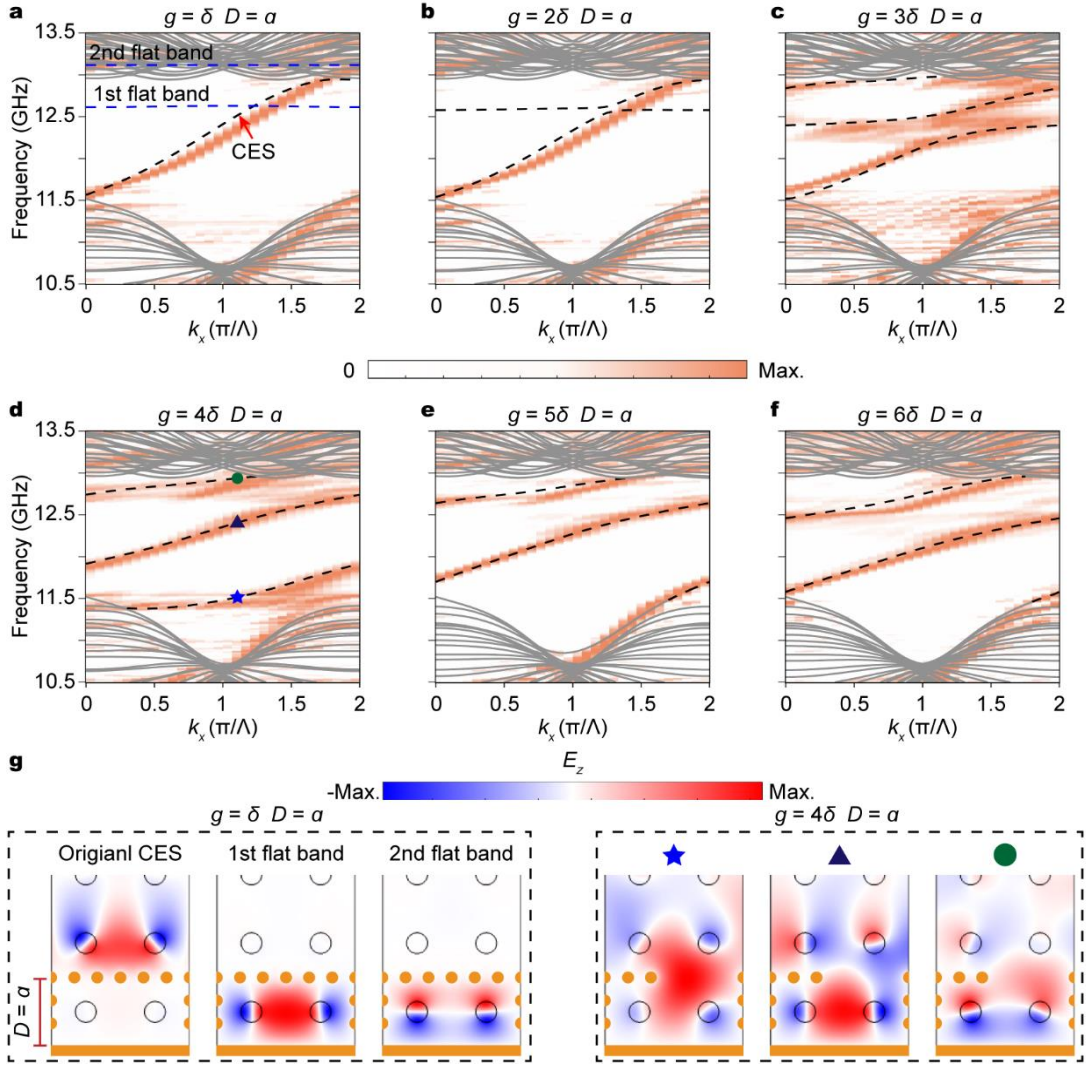

**Fig. 3 | Coupling between the CESs and the flat bands. a-f,** Experimentally measured (background colourmap) and numerically calculated dispersions (curves) for different $g$ varying from $\delta$ to $6\delta$. In all cases, $D = a$. The black and blue dashed curves denote the chiral edge modes and the flat bands, respectively. The grey curves represent projected bulk states. **g,** Simulated eigenmode profiles ($E_z$) of the edge states for the cases without (left panel) and with coupling (right panel). The corresponding frequency for each mode can be found in (**a**) and (**d**).

**Experimental observation of the broadband topological slow light through higher Brillouin zone winding.**

We then perform experiments to demonstrate the broadband topological slow light through higher momentum-space winding. In the following, we will achieve higher momentum-space winding by increasing the cavity depth $D$. The experimentally measured and numerically calculated edge dispersions for $D = 2a$, $3a$, and $4a$ are shown in Fig. 4a-c, respectively. When $D$ increases, the dispersions of the hybridised CESs wrap more times around the Brillouin zone. Since the bandwidth remains unchanged, the increases in winding numbers lead to the decreases in the group velocities, enabling the broadband topological slow light.

The higher Brillouin zone winding achieved here stems from the increased number of flat bands that are determined by the cavity size. Taking the case of $D = 4a$ as an instance (see other cases in Supplementary Information), there are four flat bands within the bandgap, which are coupled with the CESs and form new hybridised CESs, as shown in Fig. 4d. Interestingly, although these hybridised CESs exhibit completely different mode profiles, they lie on the same band. One can imagine that as the edge dispersion winds around the Brillouin zone, the edge modes evolve from one type to another.

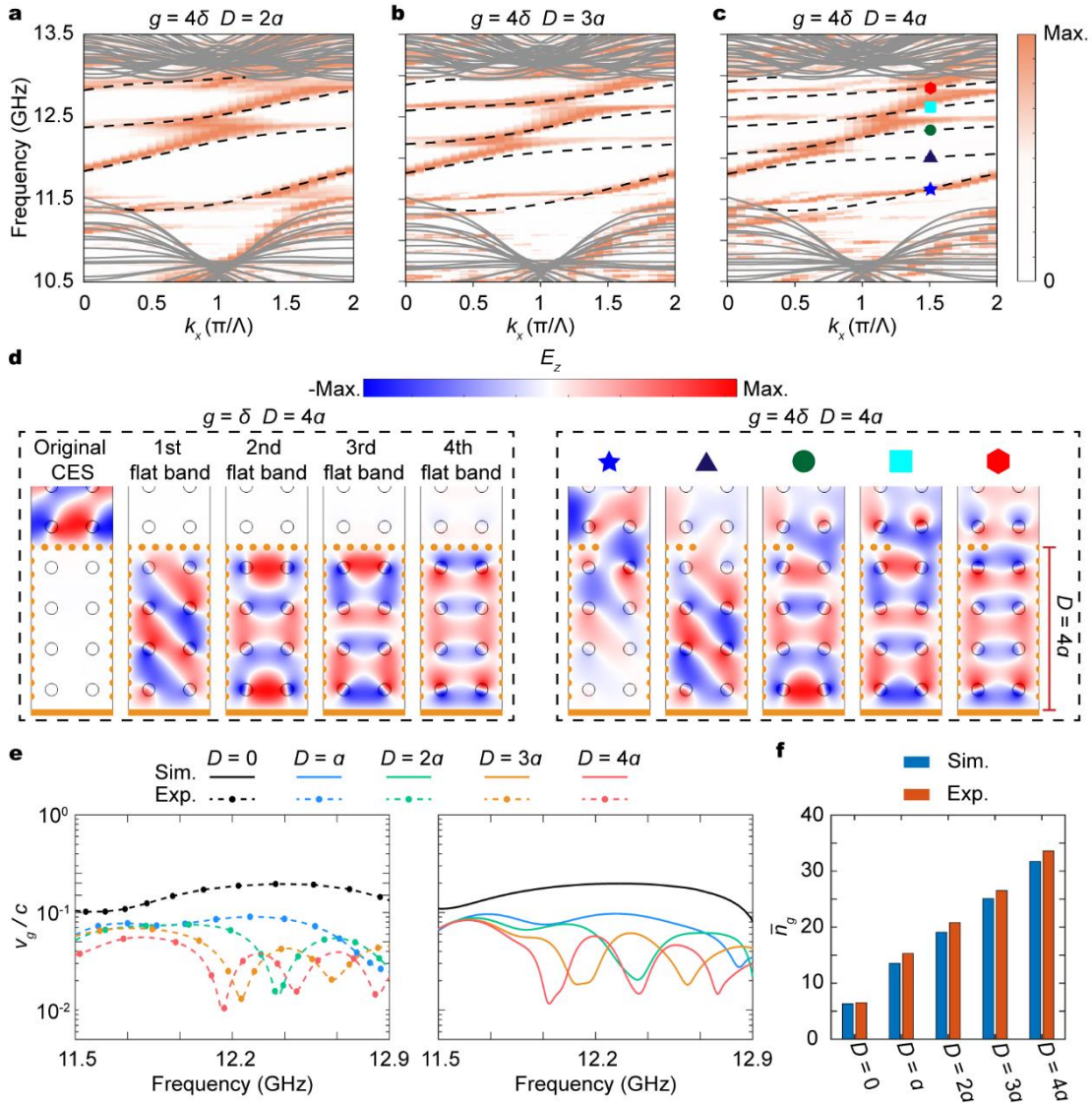

**Fig. 4 | Experimental observation of the broadband topological slow light through higher Brillouin zone winding. a-c,** Experimentally measured (colour) and numerically calculated

(curves) band structures for $D = 2a$ (**a**), $3a$ (**b**), and $4a$ (**c**), respectively. The black dashed curves are the simulated dispersions of edge modes. The solid grey curves represent projected bulk states. **d,** Simulated $E_z$-field of the eigenmode profiles of edge states without (left panel) and with (right panel) coupling. The corresponding frequency of each mode in the coupling (decoupling) regime can be found in (**c**) (Fig. S5**a** in Supplementary Information). **e,f,** Retrieved group velocities (**e**) and average group indices (**f**) from the measured and simulated dispersions of the hybridised CESs for different $D$.

To quantitatively characterise the performance of our broadband topological slow-light systems, we experimentally retrieve and numerically calculate the group velocities of the hybridised CESs, as shown in Fig. 4**e**. One can see that a higher winding number results in a significant reduction of the group velocities of the hybridised CESs for most frequencies in the topological bandgap. Remarkably, the group velocities are always positive for all frequencies within the bandgap, i.e., the dispersion curve increases monotonically as a function of the wavevector. We further calculate the average group index[43] (i.e., the slowdown factor) of the hybridised CESs within the bandgap, defined as $\overline{n}_g = \int_{f_a}^{f_b} n_g(f) df / \Delta f$, where $f_a$ = 11.5 GHz ($f_b$ =12.9 GHz) is the lower (upper) limit of the topological bandgap, $\Delta f = f_b - f_a$ is the bandgap width, $n_g(f) = c/v_g$ is the group index, and $c$ is the light speed in the vacuum. As shown in Fig. 4**f**, the measured averaged slowdown factors increase remarkably 5.1 times from 6.53 to 33.6, when $D$ varies from 0 to $4a$. The slowdown factor can be further boosted by increasing $D$ and has no strict limit in principle.

Finally, we experimentally show the robustness of the broadband topological slow light against disorder, a property that is fundamentally distinct from the conventional slow-light approaches. In experiments, we introduce two types of disorders into our system by removing several YIG rods around the edge (see Fig. 5**a,b**), and inserting a large PEC pillar into the edge (see Fig. 5**c,d**), respectively. In all cases, only the forward transmission characterised by $S_{21}$ is allowed, and the backward transmission characterised by $S_{12}$ is forbidden, within the topological bandgap. The robustness of the hybridised CESs can also be manifested from the simulated field distributions, where the hybridised CESs can pass through all kinds of obstacles and propagate unidirectionally without suffering from backscattering losses. Remarkably, we also experimentally show that in sharp contrast to the conventional slow-light systems that require cautious designs to reduce the reflection losses at the interfaces between two regions with different slowdown factors, our topological slow-light system can automatically exhibit complete transmission at those interfaces over a broad frequency range, regardless of the interface details and the group velocities at two interfaced regions (see Fig. 5**e,f**).

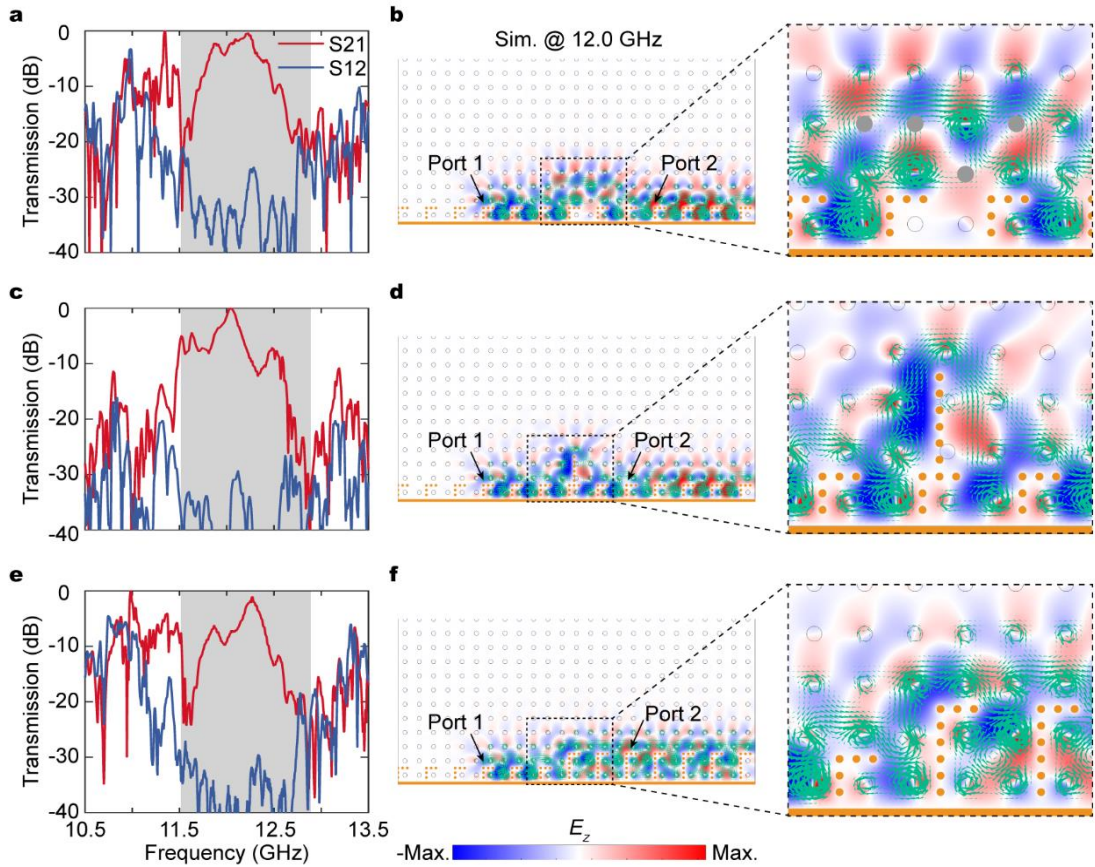

**Fig. 5 | Demonstration of the robustness of the broadband topological slow light. a,b,** Measured transmission (**a**) and simulated $E_z$-field and energy flux distributions (**b**) when four YIG pillars are removed near the edge. The grey circles in (**b**) represent removed YIG pillars. **c,d,** Measured transmission (**c**) and simulated $E_z$-field and energy flux distributions (**d**) when the edge is blocked by a large PEC obstacle. **e,f,** Measured transmissions (**e**) and simulated $E_z$-field and energy flux distributions (**f**) when the hybridised CESs pass through an interface between two regions with different group velocities. The left (right) region has $D = a$ ($D = 2a$). The $S_{21}$ ($S_{12}$) parameter is measured by placing the source antenna at port 1 (port 2) and the detector antenna at port 2 (port 1), as indicated by black arrows.

## Conclusion

We have thus experimentally demonstrated the broadband topological slow light through higher Brillouin zone winding. This is achieved by loading judiciously tailored resonances at the edge termination of the photonic topological insulator. We show that by coupling the CESs to the resonance-induced flat bands, the dispersion of the hybridised CESs can wind many times around the Brillouin zone, and the corresponding group velocities can be significantly reduced without sacrificing the operational bandwidth. Our approach, therefore, simultaneously overcomes the limitations of narrow bandwidths and susceptibility to disorders caused by fabrication imperfection. Interestingly, the newly-realised hybridised CESs closely resemble the polaritons formed by coupling photons with electric dipoles. We envision that a similar concept can be realised in nanoscale systems, where the CESs can couple with real excitations, such as phonons, excitons, magnons, and surface plasmons. The demonstration of broadband topological slow light opens up many possibilities for practical applications in enhancing light-matter interactions and nonlinearity and reducing the photonic device footprint. Finally, our approach can in principle, be extended to other classical wave systems, extending the scope of topological wave physics into slow wave engineering.

**Methods**

**Materials.** In experiments, we use samarium cobalt (SmCo) magnets with a radius of 2.2 mm and a height of 2 mm to magnetise the gyromagnetic materials. The remnant magnetic flux density of SmCo rod is measured to be 0.225 T with a deviation of ± 2.15 % along the *z*-axis. The gyromagnetic material is yttrium iron garnet (YIG), a ferrite with a measured relative permittivity 14.3 and dielectric loss tangent 0.0002, a radius of 2.2 mm, and a height of 5 mm. Its measured saturation magnetisation is $M_s$ = 1780 Gauss, and the gyromagnetic resonance loss width is 35 Oe. The magnetic flux density ($\mu_0 H_z$) at the middle plane between the two waveguide plates is about 0.043 T. When biased by an external magnetic field, the relative magnetic permeability of the YIG has a form of

$$\mu = \begin{pmatrix} \mu_r & i\kappa & 0 \\ -i\kappa & \mu_r & 0 \\ 0 & 0 & 1 \end{pmatrix}, \tag{1}$$

where $\mu_r = 1 + \dfrac{(\omega_0 + i\alpha\omega)\omega_m}{(\omega_0 + i\alpha\omega)^2 - \omega^2}$, $\kappa = \dfrac{\omega\omega_m}{(\omega_0 + i\alpha\omega)^2 - \omega^2}$, $\omega_m = \gamma\mu_0 M_s$, $\omega_0 = \gamma\mu_0 H_z$, $\mu_0 H_z = 0.043$ T is the external static magnetic flux density along $z$-axis, $\gamma = 1.76 \times 10^{11}$ s$^{-1}$T$^{-1}$ is the gyromagnetic ratio, $\alpha = 0.0088$ is the damping coefficient, and $\omega$ is the operation frequency.

**Simulations.** The dispersion relations and field patterns in Fig. 2-4, and S2-4 are numerically calculated with the eigenmode solver of the commercial software COMSOL Multiphysics. The field patterns in Fig. 5 are simulated in the frequency domain of COMSOL. To calculate the bulk band structure in Fig. 2c, Floquet periodic boundary conditions are applied in the $x$ and $y$ directions. The band structure in Fig. 3, 4, S3**a**, and S4 are calculated using a supercell that has 14 unit cells in the y direction. Floquet periodic boundary conditions are applied in the $x$ direction, and PEC boundary conditions are applied in the y direction. Both the dielectric and magnetic losses of YIG are included in numerical methods. In the frequency-domain simulation, a dipole source is placed near the interface between the photonic topological insulator and the resonator array to excite the hybridised CESs.

**Data availability**
The data that support the findings of this study are available from the corresponding author upon reasonable request.

**Code availability**
The code used in this study is available from the corresponding author upon reasonable request.

**Acknowledgments**
The work at Zhejiang University was sponsored by the National Natural Science Foundation of China (NNSFC) under Grants No. 61625502, No. 62175215, No. 11961141010, No. 61975176, No. 62071418, and No. 61931007, the Top-Notch Young Talents Program of China, the Fundamental Research Funds for the Central Universities (2021FZZX001-19), and the key Basic Research of Basic Strengthening Program of Science and Technology Commission under Grants of 2020-JCJQ-ZD-068, and 2019-JCJQ-ZD-128-6.

**Author contributions**
Y.Y., and H.X. initiated the idea. Y.Y., F.C., and P.Z. designed the structures and the experiments. F.C., P.Z., Y.P., M.W., and Y.H. conducted the experiments. F.C., Y.Y., H.X., B.Z., P.Z., L.Z., and Q.C. did the theoretical analysis. F.C., Y.Y., B.Z., P.Z., W.Y., and H.C. wrote the manuscript and interpreted the results. Y.Y., P.Z., W.Y., H.C., and

B.Z. supervised the project. All authors participated in discussions and reviewed the manuscript.

## Competing interests
The authors declare no competing interests.

## Additional information
**Supplementary information** is available for this paper.

**Correspondence and requests for materials** should be addressed to Yihao Yang, or Hongsheng Chen.